\documentclass[twocolumn,showpacs,preprintnumbers,amsmath,amssymb,superscriptaddress]{revtex4}
\usepackage{graphicx}
\usepackage{dcolumn}
\usepackage{bm}
\usepackage{longtable}

\newcommand{\bea}{\begin{eqnarray}}
\newcommand{\eea}{\end{eqnarray}}
\newcommand{\be}{\begin{equation}}
\newcommand{\ee}{\end{equation}}

\usepackage{amsfonts}
\usepackage{amssymb}
\usepackage{amsmath}
\usepackage{amsthm}
\usepackage{graphicx}

\begin{document}


\title{Slow Encounters of Particle Pairs in Branched Structures}

\author{Elena Agliari}
\affiliation{Dipartimento di Fisica, Sapienza Universit\`a di Roma}
\affiliation{INdAM, Gruppo Collegato di ``Tor Vergata'', Roma}
\author{Alexander Blumen}
\affiliation{Theoretische Polymerphysik, Universit\"{a}t Freiburg}
\author{Davide Cassi}
\affiliation{Dipartimento di Fisica, Universit\`a di Parma}

\date{\today}


\begin{abstract}
On infinite homogeneous structures, two random walkers meet with certainty if and only if
the structure is recurrent, i.e., a single random walker returns to its starting point with probability $1$. However, on general inhomogeneous structures this property does not hold and, although a single random walker will certainly return to its starting point, two moving particles may never meet. This striking property has been shown to hold, for instance, on infinite combs.
Due to the huge variety of natural phenomena which can be modeled in terms of encounters between two (or more) particles diffusing in comb-like structures, it is fundamental to investigate if and, if so, to what extent similar effects may take place in finite structures. 
By means of numerical simulations we evidence that, indeed, even on finite structures, the topological inhomogeneity can qualitatively affect the two-particle problem. In particular, the mean encounter time can be polynomially larger than the time expected from the related one particle problem.  
\end{abstract}

\maketitle

\section{Introduction}
Network theory generally refers to the investigation of graphs (meant as a representation of a set of discrete objects in mutual interaction with each other), focusing on their topological properties, as well as on the dynamics of arbitrary agents spreading on them. 
In particular, diffusion processes occurring on complex networks (e.g. lacking translational invariance) can give rise to anomalous behaviors strongly related to the underlying topology \cite{BC-JPA2005,ben-2000,Benichou-Nature2007}.

In the last decade network theory has attracted an increasing interest and an impressive number of results, analytical and/or numerical, is nowadays available.
Most of them are concerned with very popular models, like scale-free networks, random graphs \`a la Erd\"os-R\'enyi, small-world networks, transfractals \cite{Newman-2010,Rozenfeld-NJP2007,Meyer-PRE2012}. These models have proved to be very effective in describing \emph{superstructures}, namely \emph{artificial structures} such as the World Web Web, Internet, social networks, etc. 
On the other hand, when dealing with \emph{natural structures}, such as macromolecules, disordered materials, biological systems, the previous models are no longer so adequate since geometries generally occurring in Nature are typically embeddable in low-dimensional spaces (this also means that their degree is finite) and often have a tree-like architecture (see e.g., \cite{Vekshin-1997,GB-AdPolSc2005,Frauenrath-ProgPolymSci2005,Thiriet-2013}). 

%
 
A very versatile and interesting model for such structures is given by combs, which, as we are going to explain, can strongly affect the underlying dynamic processes.

As a paradigmatic example, here we focus the attention on reaction-diffusion processes, namely systems where a given event is triggered as two or more diffusive particles happen to be sufficiently close.
There exist many basic phenomena which can be modeled in these terms and which stem from different fields, such as pharmacokinetics  (where the branched topology of the circulatory systems \cite{Alberts-2002,Welte-CurrBio2004,Santamaria-Neur2006,Arkhincheev-JM2011,Thiriet-2013}) is known to deeply affect the diffusion of drugs \cite{Marsh-QMed2008}), chemical physics (where energy transfer in comb polymers  \cite{Casassa-JPolymSci1966,Douglas-Macromol1990} and dendronized polymers \cite{Frauenrath-ProgPolymSci2005} can exhibit anomalous diffusion \cite{Vekshin-1997}), in neuroscience (where the properties of calcium transport and reaction in spiny dendrites \cite{Yuste-MIT2010,Nimchinsky-AnnRevPhys2002,Santamaria-Neur2006} can be related to neural plasticity \cite{Mendez-2010,Iomin-PRE2013}), in condensed matter (where combs serve as a model for porous materials \cite{Weiss-PhilMag1987,Arkhincheev-JM2011, Arkhincheev-IEEE2013,Stanely-PRB1984,Tarasenko-MMM2012}) and even in architecture (where optimal diffusion through ecological \cite{Hannuen-EcoMod2002} as well as urbanistic \cite{Medina} systems is envisaged).

Recently, the problem of two simple random walkers moving on a regular, infinite comb has been rigorously analyzed \cite{Peres-ECP2004,Chen-EJP2011}, showing very interesting phenomena: different from homogeneous structures where the two-particle problem (i.e. the problem of finding out how likely is that two particles eventually meet) can be mapped into a one-particle problem (i.e. the problem of finding out how likely is that one particle eventually reaches a given fixed target), in combs the two problems are not only intrinsically distinct but, also, their solution are strikingly different. In fact, a single particle randomly moving on a comb is certain to eventually visit any site, while two particles display a finite probability of never encountering each other, notwithstanding their initial position.
This result has been rigorously proven for infinite combs and suggests that the topological inhomogeneities of such structures may lead to dramatic effects for reaction-diffusion processes. However, as real phenomena necessarily occur in \emph{finite} structures, it is fundamental to investigate if and, if so, to what extent similar effects may take place in finite structures \cite{Agliari-2014}. 

Indeed, in finite structures,  we expect that at intermediate times (i.e. times long enough to see the emergence of asymptotic behaviors, but not too long for the random walk to realize the finiteness of the substrate) the two-particle problem will exhibit non-trivial features. 

In the following we analyze the two-particle problem on different kinds of finite branched structures, here generically referred to as $\mathcal{G}$, and we will focus on the probability distribution $P_{\mathcal{G}}(t, L)$ for the time $t$ to first meet on a structure of size $L$. From this quantity we can derive the related moments and, in particular, the mean first encounter time $\tau_{\mathcal{G}}(L)$. Interestingly, as we will show, $P_{\mathcal{G}}(t,L)$ may display extremal points mirroring the existence of characteristic time scales. 
Moreover we find that, according to initialization, $\tau_{\mathcal{G}}$ may scale ``anomalously'' with $L$, or, more precisely, that the mean encounter time can be polynomially larger than the time expected from the related one particle problem.  

In fact, in order to better highlight the peculiarity of such results, we also consider the case of reactions between two particles, being one mobile and the other immobile. Again, we are interested in the time for the reaction to (first) occur and we measure the related probability distribution $Q_{\mathcal{G}}(t, L)$ and its average value $\zeta_{\mathcal{G}}$.

\section{Dynamics on finite combs}
In this section we distinguish between the case of regular combs, referred to as $\mathcal{C}$ (see Fig.~
$1$,
left panel), and the case of irregular combs, referred to as $\mathcal{B}$, where the length of the side chains is random (see Fig.~
$1$,
 right panel).

 \begin{figure}\label{fig:Combs}
\includegraphics[width=4.25cm]{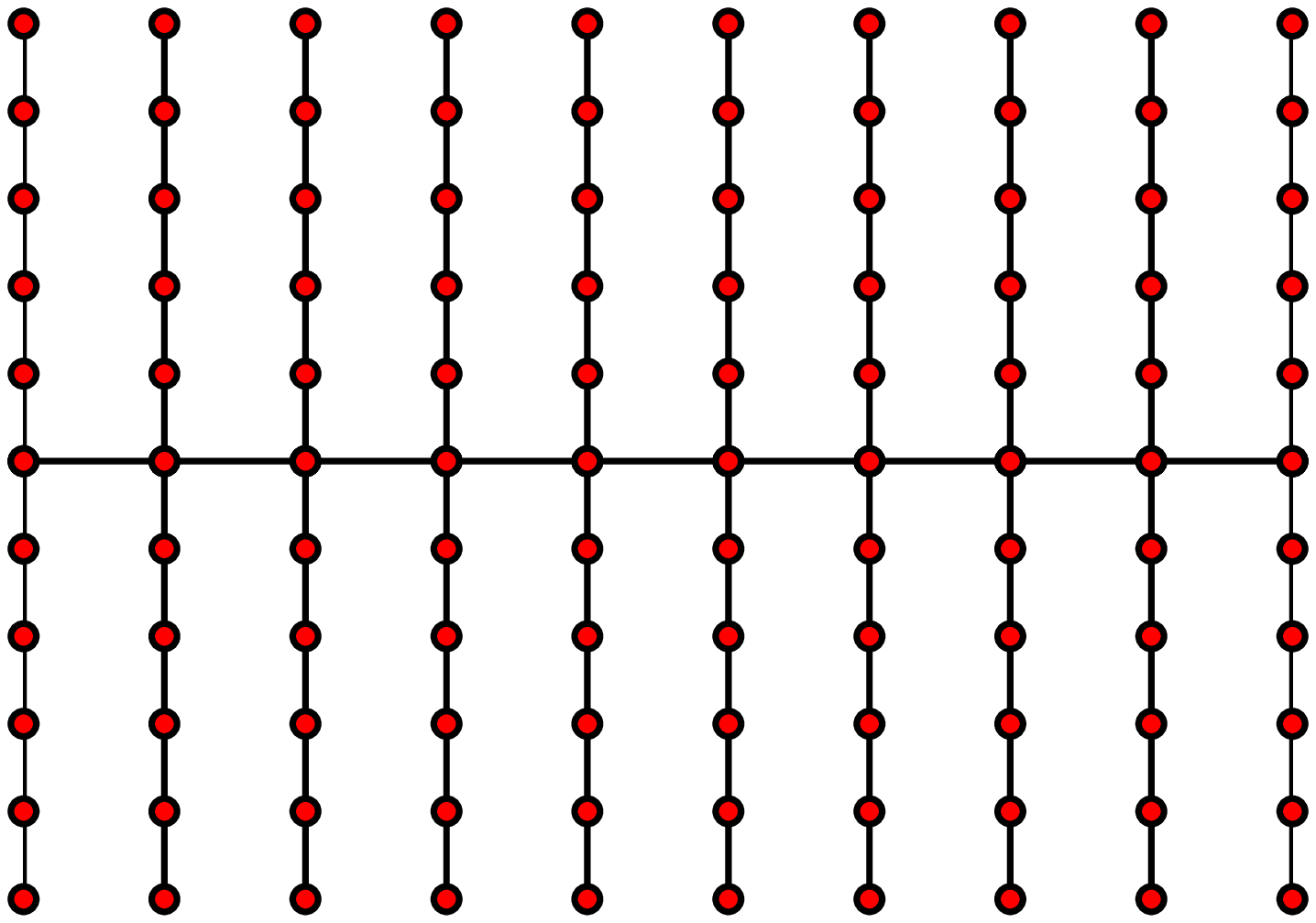}
\includegraphics[width=4.25cm]{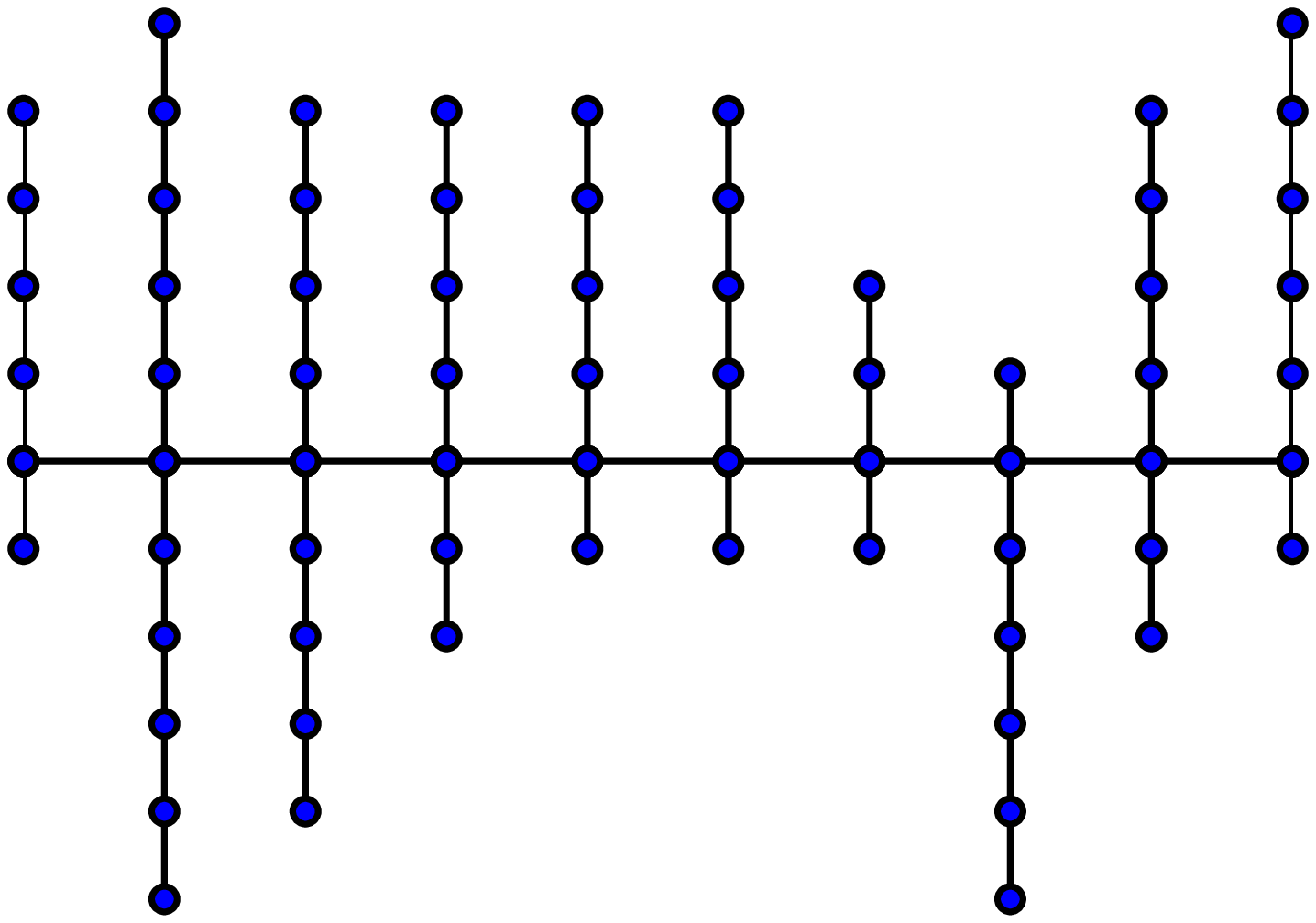} 
 \caption{Examples of a regular comb lattice with $L_x=10$ and $L_y=10$ (left panel) and of a random comb lattice with $L_x=10$ and side chains of length randomly drawn from a uniform distribution with average $3$ (right panel).}
 \end{figure}

  \begin{figure}\label{fig:DISTR1}
\includegraphics[width=7cm]{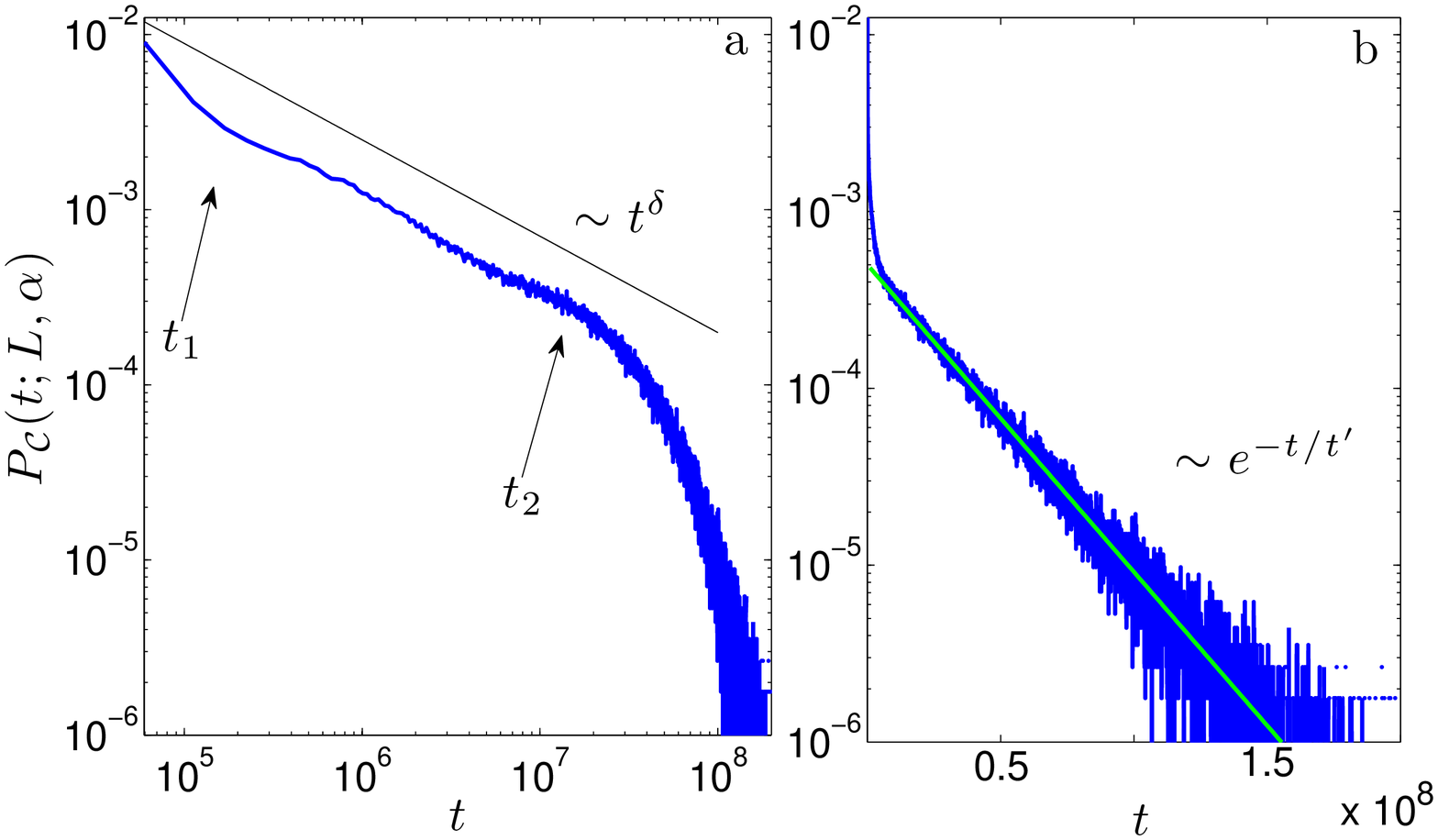}
\includegraphics[width=7cm]{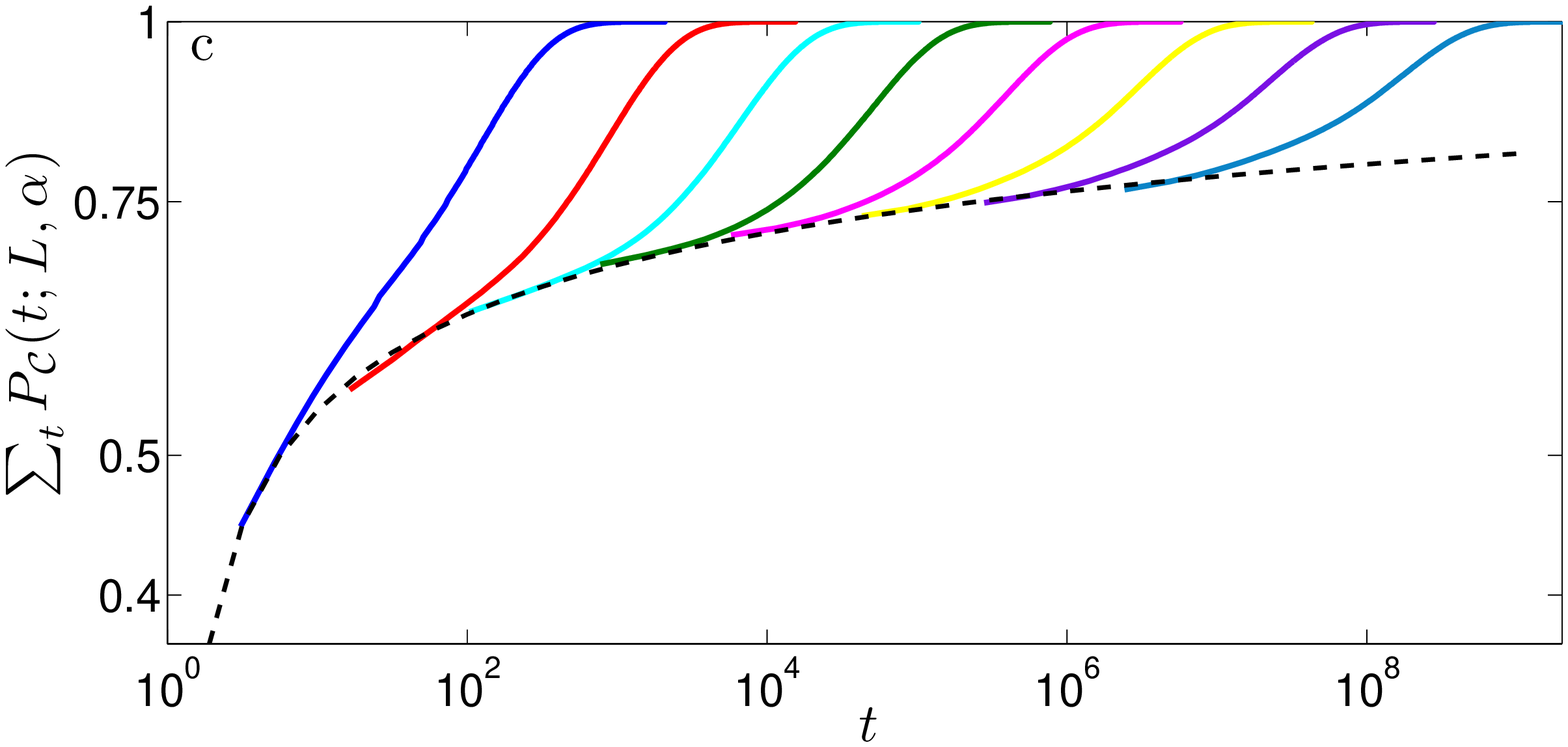} 
 \caption{Probability distribution $P_{\mathcal{C}}(t;L,\alpha)$ for the first-encounter time on a deterministic comb of linear sizes $L_x=256$ and $L_y=4L_x$, plotted on a log-log scale (panel $a$) to highlight intermediate times (in between $t_1$ and $t_2$) and on a semi-logarithmic scale (panel $b$) to highlight long times (larger than $t_2$). The best fits highlighted correspond to $\delta = -0.55$ and to $t^{\prime} = (2.5 \pm 0.1) \times 10^7$. The values of the characteristic times $t_1$ and $t_2$ pertaining to different sizes are shown in Fig.~$3$. The cumulative distribution $\sum_t P_{\mathcal{C}}(t,L,\alpha)$ (panel $c$) is shown for different choices of $L$:  the curves from left to right correspond to $L=2^{k}$, with $k=2, ..., 9$, respectively. The envelope of the related starting points is fitted by the curve $y = -a/ \sqrt{\log x} +b$, with $a = -0.63 \pm 0.03$ and $b = 0.92 \pm 0.02$. The data shown in these panels have been obtained via numerical simulations and the sample is made of $10^7$ realizations for every $L$. }
 \end{figure}
 
\subsection{Regular Combs} \label{sec:Reg}
Regular combs are built by fixing the length $L_x$ (for simplicity $L_x$ is even) of the backbone and by attaching to each of its sites two side chains of length $L_y/2$, where $L_y = \alpha L_x$, being $\alpha \in \mathbb{N}$; in this way the overall number of sites $N$ is $L_x(\alpha L_x+1)$. To simplify notation hereafter $L_x$ will be referred to as $L$.
Periodic boundary conditions are applied to the backbone, while reflecting boundary conditions are applied to the side chains. 
This kind of structure can be embedded in the two dimensions ($d=2$) and extensions to higher dimensional spaces ($d=3, 4, ...$) can also be realized (see also Sec.~\ref{sec:Higher}).
Regular, infinite combs have been extensively analyzed in \cite{CR-MPL1992}, where it was shown that the spectral dimension $\tilde{d}$ is given by $\tilde{d} = 2 (1-2^{-d})$. We recall that the latter provides information about the dynamic properties of the graph, for instance, the probability for a random walker to return to its starting point scales asymptotically like $\sim t^{-\tilde{d}/2}$, (see e.g., \cite{BC-JPA2005}).

\quad

\emph{Particles starting from the same initial position}\\
Let us consider two random walkers initially placed on the same site of the backbone. The walkers are allowed to move up to time $t$, when they again occupy the same site for the very first time. The probability distributions $P_{\mathcal{C}}(t ; L, \alpha)$ obtained from numerical simulations are shown in Fig.~$2$.

Interestingly, $P_{\mathcal{C}}(t; L,\alpha)$ displays three different regimes, distinguished by two ``critical points'' corresponding to two characteristic time scales which we denote by $t_1$ and $t_2$, respectively.
More precisely, at intermediate times, i.e. $t_1 < t < t_2$, the probability distribution decays as a power law, as expected for an infinite structure (e.g., see \cite{CC-PRE2012}), suggesting that this time range corresponds to the asymptotic regime; on the other hand, at long times,  i.e. for $t>t_2$, the probability distribution decays exponentially, suggesting that this time range corresponds to the emergence of finite size effects which provide a ``boost'' in the likelihood for the two particles to meet.
We stress that the heavy-tailed distribution means that the first encounter time is broadly spread with a large (indeed infinite in the thermodynamic limit) mean, as expected due to the finite collision property displayed by such structures \footnote{We say that a graph has the \emph{finite} (\emph{infinite}) \emph{collision property} if two independent random walks on it, starting from the same node, meet finitely (infinitely) many times almost surely. See, e.g., \cite{CC-PRE2012}.}; in particular, by fitting the data we find $P_{\mathcal{C}}(t; L,\alpha) \sim t^{\delta}$, with $\delta \approx -0.55$. 

The points $t_1$ and $t_2$ can therefore be extracted for different sizes $L$ as the onset of a power law behavior and of an exponential behavior, respectively. These values are shown in Fig.~$3$, where we evidence that $t_1$ scales like $t_1 \sim L_x L_y$, while $t_2$ scales like $t_2 \sim L_x L_y^{\gamma}$, where $\gamma \approx 1.75$; as we will show in the following, $t_2$ is closely related to the mean first encounter time $\tau_{\mathcal{C}}$.

 \begin{figure}\label{fig:CRIT1}
\includegraphics[width=7cm]{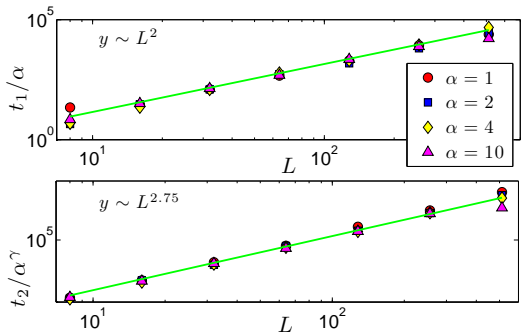}
 \caption{From the distribution $P_{\mathcal{C}}(t;L,\alpha)$ we extracted $t_1(L,\alpha)$ (upper panel) and $t_2(L,\alpha)$ (lower panel). These characteristic times were divided by $\alpha$ and by $\alpha^{\gamma}$, respectively, and plotted versus $L$. By assuming $\gamma=1.75$ we obtain a nice collapse of data points. Best fits (solid lines) correspond to power laws with exponents $2$ and $\gamma$, respectively, hence suggesting $t_1 \sim L_x L_y$ and $t_2 \sim L_x L_y^{\gamma}$.}
 \end{figure}
 
In order to highlight the size dependence of the first-encounter time distribution it is convenient to consider the cumulative distribution $\sum_t P_{\mathcal{C}}(t; L,\alpha)$. As shown in Fig.~2, panel $c$, at short times the cumulative distributions pertaining to different values of $\alpha$ overlap nicely with the curve expected from the infinite-structure case $P_{\mathcal{C}}(t)$; otherwise stated, as long as $t< t_1(L,\alpha)$, $P_{\mathcal{C}}(t; L, \alpha)$ is indistinguishable from $P_{\mathcal{C}}(t)$, consistent with the fact that finite size behavior has not emerged yet. Thus, by fitting the early-time envelopes of $P_{\mathcal{C}}(t; L, \alpha)$ we get an estimate for $P_{\mathcal{C}}(t)$, which turns out to saturate to $1$ with a rate scaling as $1/\sqrt{\log t}$.

Now, from the distribution $P_{\mathcal{C}}(t ; L,\alpha)$, one can derive the mean first encounter time
\be \label{eq:tc}
\tau_{\mathcal{C}} (L,\alpha) \equiv \sum_{t=0}^{\infty} t \, P_{\mathcal{C}}(t ; L,\alpha).
\ee
Results for different values of $\alpha$ and $L$ are shown in Fig.~$4$.
By properly fitting the data we find that
\be \label{eq:tau}
\tau_{\mathcal{C}} (L,\alpha) \sim L_x L_y^{\gamma} \sim L^{1 + \gamma}.
\ee
Interestingly, we can speculate that $1 + \gamma = d + \tilde{d}/2 \approx 2.75$.

\begin{figure}\label{fig:TAU1}
\includegraphics[width=7cm]{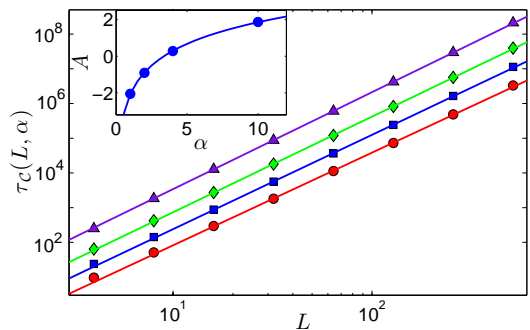}
 \caption{Mean first encounter time $\tau_{\mathcal{C}}(L, \alpha)$ versus $L$ for different choices of $\alpha$, represented by different symbols, so that from below $\alpha=1, 2, 4, 10$. 
 In a logarithmic scale these data are well fitted by a linear law (solid line), i.e. $\log(\tau_{\mathcal{C}}) = A + \gamma \log(L)$, where $\gamma$ turns out to be independent of $\alpha$, while $A$ depends logarithmically on $\alpha$ (see the inset), hence suggesting for $\tau_{\mathcal{C}}(L, \alpha)$ the overall behavior given by Eq.~\ref{eq:tau}.}
 \end{figure}

\quad

Let us now consider the case where one of the two particles is immobile and fixed at a given point on the backbone, while the other is allowed to perform a random walk starting from the same site. Again, we are interested in the time the particles meet for the first time; in this case, the reaction time corresponds to the first return time of the mobile particle.

Results for the distribution $Q_{\mathcal{C}}(t;L,\alpha)$ of the first-encounter time and for the related cumulative distribution $\sum_t Q_{\mathcal{C}}(t;L,\alpha)$ are shown in Fig.~$5$. 
Analogously to $P_{\mathcal{C}}(t;L,\alpha)$, we can distinguish an early-time regime, an intermediate regime, and a late-time regime.
The second one is the most interesting; it displays a power-law decay with the probability distributions scaling as $t^{\rho}$, with $\rho \approx - 1.25$. Notice that $\rho < \delta$, namely, when a particle is fixed, the distribution is less broad, consistent with the fact that, when both particles are mobile, the reaction much less likely due to the finite collision property. For large sizes $L$ the late-time regime exhibits a peak corresponding to the mobile particle being close to the starting point. 

\begin{figure}\label{fig:DISTR1b}
\includegraphics[height=4.5cm]{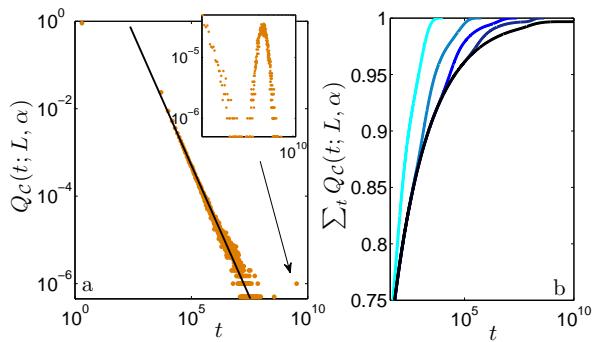}
\caption{Probability distribution $Q_{\mathcal{C}}(t;L,\alpha)$ for the time of first return to the starting point on the backbone for a random walker placed on a regular, square comb of linear size $L_x=4096$ (panel $a$). Different choices for the backbone size are compared by considering the cumulative distribution $\sum_t Q_{\mathcal{C}}(t,L,\alpha)$ (panel $b$): from left to right the curves correspond to $L= 4^k$, with $k=2,3,4,5,7$, respectively. Notice that only the mid-to-long time regime is shown: curves overlap up to a characteristic time after which finite size effects emerge. The data shown here have been obtained via numerical simulations and, for every $L$, the sample contains $10^7$ realizations.}
 \end{figure}
 
As for the mean encounter time
\be
\zeta_{\mathcal{C}} (L, \alpha) \equiv \sum_{t} t \, Q_{\mathcal{C}}(t;L,\alpha),
\ee
 we get  (see Fig.~$6$)
 \be \label{eq:zeta}
 \zeta_{\mathcal{C}} (L, \alpha) \sim L_x L_y  \sim L^2.
 \ee

This result can be understood by mapping the random walk on the comb into a continuous-time random walk on a linear chain, where the waiting time distribution is identical for all nodes and has an average given by the mean time spent by the original walk on the side-chain; this mean waiting time ultimately corresponds to the mean time $\tau_1$ spent by a random walk, which started on the origin of a finite chain of length $L_y$, to first return to its initial point \cite{Matan-JPA1989, Redner-2001book,SYRL-PRE2009}. Then, denoting with $\tau_2$ the mean number of steps taken by a continuous-time random walk to first return to its starting point on a ring of length $L_x$, we can derive $\zeta_{\mathcal{C}} \sim \tau_1 \tau_2$. Now, recalling that $\tau_1$ and $\tau_2$ scale linearly with the size of the underlying structure (see e.g., \cite{Redner-2001book}), we finally get $\zeta_{\mathcal{C}} \sim \tau_1 \tau_2 \sim L_y L_x$, as anticipated.

\begin{figure}\label{fig:TAU1b}
\includegraphics[width=7cm]{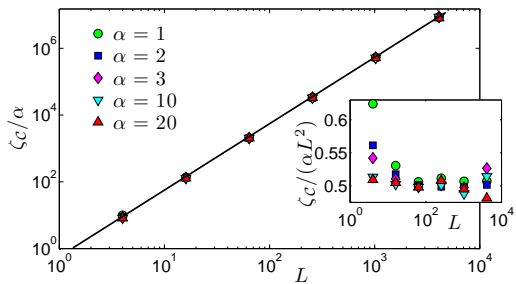}
 \caption{Mean return time for a particle starting on the backbone of a regular comb of linear size $L$ and side chains of length $\alpha L$; several choices of $\alpha$ are considered, as given in the legend. In order to get data collapse, $\zeta_{\mathcal{C}} (L, \alpha)$ is divided by $\alpha$. The data from the simulations (symbols) are best fitted by the power law $y = L^2/2$ (solid line), hence suggesting the overall behavior given by Eq.~\ref{eq:zeta}. The inset highlights, in a linear scale, the scatter of the data around the approximation $\alpha L^2$.}
 \end{figure}

Notice that the two-particle problem and the one-particle problem lead to qualitatively different results, having $\tau_{\mathcal{C}}/\zeta_{\mathcal{C}} \sim L^{\gamma - 1} \rightarrow \infty$.

\quad

\emph{Particles starting from different initial positions on the backbone}\\
Let us consider two random walkers initially placed on two distinct sites of the backbone; to fix the ideas let us choose two nodes at the maximal mutual distance $L/2$ (of course the parity of the two starting nodes has to be the same). The walkers are allowed to move up to time $t$, where they occupy the same site for the very first time. The probability distributions $P^{\prime}_{\mathcal{C}} (t;L, \alpha)$ obtained from numerical simulations are shown in Fig.~$7$
(left panel).

Such distributions peak at a point $t_1$ which defines a characteristic time scale for the encounter to occur.
We extracted $t_1$ for different values of $L$ and of $\alpha$: the results are summarized in Fig.~$7$ 
(right panel).
The data collapse when divided by $\alpha^{\gamma^{\prime}}$, with $\gamma^{\prime} \approx \gamma$; such collapsed data can be fitted by $L^3$, in such a way that we get the overall behavior $t_1 \sim L_x^{3 - \gamma^{\prime}} L_y^{\gamma^{\prime}}$.

 Again, the ``critical'' points of the distributions are intimately related to the mean encounter time
 \be
 \tau_{\mathcal{C}}^{\prime}(L, \alpha) \equiv \sum_{t} t  P^{\prime}_{\mathcal{C}} (t,L).
 \ee
 In fact, as shown in Fig.~$8$,
 the mean encounter time scales as $t_1$; we have namely
 \be  \label{eq:tauc2}
 \tau_{\mathcal{C}}^{\prime} (L, \alpha) \sim L_x^{3 - \gamma^{\prime}} L_y^{\gamma^{\prime}} \sim L^3.
 \ee

Notice that when both particles are mobile their (extensive) initial distance enters sublinearly into the mean encounter time, since $ \tau_{\mathcal{C}}^{\prime} /  \tau_{\mathcal{C}} \sim L^{2 - \gamma}$.

\begin{figure}\label{fig:DISTR2a}
\includegraphics[width=7cm]{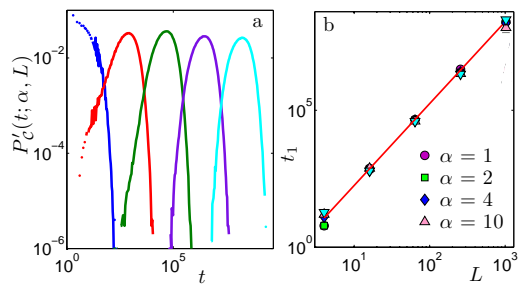}
 \caption{Distribution $P^{\prime}_{\mathcal{C}} (t,L)$ for the first encounter time of two particles starting on the backbone at a distance $L/2$ of each other. Different choices for the backbone length are considered and compared: from left to right $L_x = 4^k$, with $k=1,2,...,5$; here $L_x = 2 L_y$, that is the sides of the comb form a square. On the right panel we show the fit for the extremal time $t_1$ divided by $\alpha^{\gamma}$ in order to get data collapse.
The data shown here have been obtained via numerical simulations and, for every $L$, the sample contains $10^7$ realizations.}
 \end{figure}
 
\begin{figure}\label{fig:TAU2a}
\includegraphics[width=7cm]{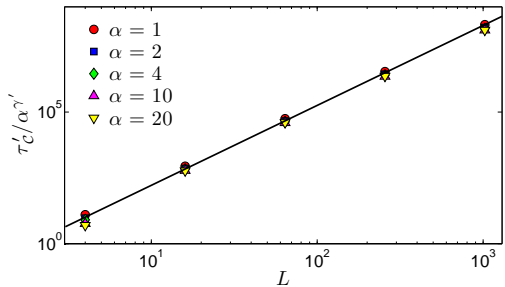}
 \caption{Mean encounter time $\tau^{\prime}_{\mathcal{C}}(L,\alpha)$ for two moving particles which start from nodes on the backbone at a distance $L/2$. The collapse of the data points is obtained by dividing $\tau^{\prime}_{\mathcal{C}}(L,\alpha)$ by $\alpha^{\gamma^{\prime}}$ and assuming that $\gamma^{\prime} = 1.75$. Numerically obtained data (symbols) are best fitted by a power law with exponent $3$ (solid line), hence suggesting the overall behavior given by Eq.~\ref{eq:tauc2}.}
 \end{figure}
 
\quad

Finally, we investigate the case of one immobile particle, fixed at a given site on the backbone, and one mobile particle starting at a distance $L/2$ on the backbone and performing a random walk until it reaches for the first time the fixed particle.

Results for the distribution of the first encounter time $Q^{\prime}_{\mathcal{C}}(t ; L,\alpha)$ are shown in Fig.~$9$ (left panel).
Analogously to $P^{\prime}_{\mathcal{C}}(t ; L,\alpha)$, namely to the case of two mobile particles, $Q^{\prime}_{\mathcal{C}}(t ; L,\alpha)$ peaks at a characteristic time denoted by $t_1$. We extracted the value of $t_1$ for different choices of the comb sizes and summarized the results in Fig.~$9$ (right panel): the overall behavior is given by  $t_1 \sim L_y L_x^2$. Again, $t_1$ defines a characteristic time scale for the encounter to occur and its behavior is mirrored by the mean encounter time 
\be
\zeta^{\prime}_{\mathcal{C}}(L, \alpha) \equiv \sum_t t \, Q^{\prime}_{\mathcal{C}}(t ; L,\alpha).
\ee 
In fact, for the latter we found
\be  \label{eq:zetac2}
\zeta^{\prime}_{\mathcal{C}}(L, \alpha) \sim L_y L_x^2 \sim L^3,
\ee
as shown in Fig.~$10$.

\begin{figure}\label{fig:DISTR2b}
\includegraphics[width=7cm]{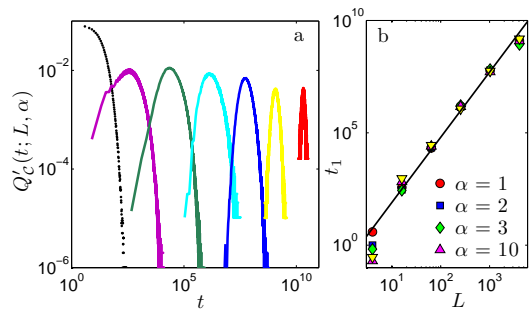}
 \caption{Panel $a$: Distribution $Q^{\prime}_{\mathcal{C}}(t ; L,\alpha)$ for the time to first reach a point on the backbone distant by $L/2$. Different choices for the backbone are considered and compared: from left to right, the curves refer to $L_x=4^{k}$, with $k=1,2,...,7$. In every case $L_x = 2 L_y$, i.e. the sides of the comb form a square. Panel $b$: The values of the characteristic time $t_1$ (symbols) corresponding to the maximum of the distribution $Q^{\prime}_{\mathcal{C}}(t ; L,\alpha)$ were extracted for different choices of $\alpha$ (see the legend) and plotted versus $L$. Notice that the data collapse is obtained dividing $t_1$ by $\alpha$. The best fit (solid line) corresponds to a power law with exponent $3$. The data shown in these panels have been obtained via numerical simulations and, for every $L$, the sample contains $10^7$ realizations. }
 \end{figure}

Such a scaling can be understood by mapping the problem into a continuous-time random walk picture as done before for $\zeta_{\mathcal{C}}$. The extra factor $L$ appearing in Eq.~\ref{eq:zetac2} is due to the mean number of steps needed by the walker to attain for the first time a distance $L/2$ along the backbone, which scales as $L^2$. In fact, the number of steps required to cover a distance $x\sim L/2$ on a ring of length $L$ scales as $x(L-x) \sim L^2$ \cite{Redner-2001book}.

We also notice that in this case (at least for the small sizes considered) there is no qualitative difference in the mean reaction time according to whether one of the two particles is kept fixed or not, that is, $\tau^{\prime}_{\mathcal{C}} / \zeta^{\prime}_{\mathcal{C}} \sim 1$. Moreover, in both cases, the mean time scales super-linearly with the total volume, namely with $N^{3/2}$.

In conclusion, we expect that the leading scaling with $L^3$ represents an upper bound for the mean time to encounter of two particles started at any mutual distance. In particular, we verified that when particles start at the maximal distance $2L_y + L_x/2$ (namely on extremal points of farthest teeth), the mean encounter time scales as $L^3$. In fact, the time to reach the backbone contributes with a sub-leading term $L$. Moreover, we checked that when the starting sites are chosen randomly, the mean encounter time (where the average runs now also over the initial positions) scales like $L^3$.

\subsection{Higher-order branched structures} \label{sec:Higher}
Many natural branched structures, such as neurites and dendrites, may display so-called secondary, tertiary and higher-order branches (see e.g., \cite{Jan-NatNeur2010}), that is to say, the (first-order) side-chains of a comb can be further branched by (second-order) side-chains, and so on in a recursive way.
Combs exhibiting $(d-1)$-th order branches can be embedded in a $d$-dimensional space and are therefore called $d$-dimensional combs \cite{CR-MPL1992,CR-PRL1995}, hereafter referred to as $\mathcal{C}_d$. One can therefore ask whether the slowing down phenomena evidenced for $2$-dimensional combs also emerge in higher-dimensional combs. 
 
Here, we consider the mean encounter time $\tau_{\mathcal{C}_d}$ for two random walkers starting on the same site on the backbone and the mean return time $\zeta_{\mathcal{C}_d}$ for a random walker starting on the backbone, focusing on their dependence on the linear size of the structure; for simplicity we restrict ourselves to the case of combs with the same linear size along all directions, i.e. $L_x=L_y=L_z \equiv L$, for $d=3$ and analogously for higher orders.

Results are summarized in Fig.~$11$:
as highlighted by fitting functions, $\zeta_{\mathcal{C}_d}$ follows the behavior expected for Euclidean structures\footnote{We recall that on Euclidean structures such as (hyper) cubic lattices  the mean time for two particles to meet for the first time scales linearly with the volume of the underlying lattice and the same scaling holds for the mean return time.},while $\tau_{\mathcal{C}_d}$ grows qualitatively faster, namely
\begin{eqnarray}
\tau_{\mathcal{C}_d} &\sim& L^{1+ \gamma_d},\\
\zeta_{\mathcal{C}_d} &\sim& L^d,
\end{eqnarray}
where, $1 + \gamma_3 \approx 3.88$ and $1 + \gamma_4 \approx 4.94$, hence confirming the relation $1 + \gamma_d = d + \tilde{d}/2 = d+1 -2^{-d}$, proposed in Sec.~\ref{sec:Reg}.

Thus, even for higher-order combs the two-particles encounter turns out to be slow and $\tau_{\mathcal{C}_d} / \zeta_{\mathcal{C}_d} \sim L^{1-2^{-d}}$. Moreover, for $d \gg 1$, we expect that  the mean encounter time scales as $\sim L^{1+ d} = V^{1 + 1/d} \sim V$.

\begin{figure}\label{fig:comb_d}
\includegraphics[width=8cm]{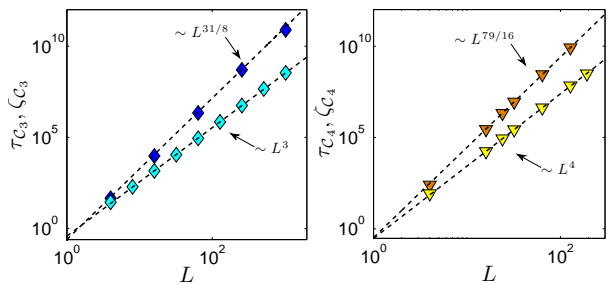}
 \caption{Mean encounter times $\tau_{\mathcal{C}_d}$ (darker) and $\zeta_{\mathcal{C}_d}$ (brighter) with fitting functions for $d=3$ (panel $a$) and for $d=4$ (panel $b$). 
The best fits (dashed lines) correspond to power laws, as reported. The data shown in these panels have been obtained via numerical simulations and, for every $L$, the sample contains $10^7$ realizations.}
 \end{figure}

\subsection{Randomly branched structures}
In this section we present results obtained for branched structures, which differ from those analyzed before, by exhibiting some degree of randomness. More precisely, we will consider structures with a backbone of length $L_x$ and side chains whose length $L_y$ is a stochastic variable with a given (finite) mean value $\langle L_y \rangle$ (see Fig.~$1$, right panel).
These models are closer to biological structures; for instance, when considering transport processes in spiny dendrites one finds that the distribution of spines along the dendrite, their sizes and shapes appear to be highly random \cite{Nimchinsky-AnnRevPhys2002,Mendez-CSF2013}.

\begin{figure}\label{fig:TAU2b}
\includegraphics[width=6cm]{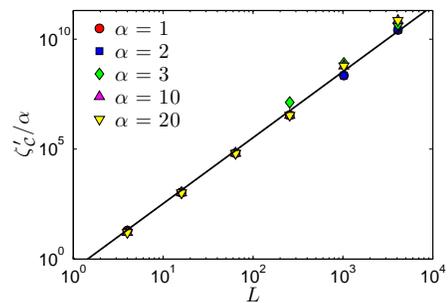}
\caption{Mean encounter time divided by $\alpha$, namely $\zeta^{\prime}_{\mathcal{C}}/\alpha$ when one particle is immobile and the other is mobile starting from a site at a distance $L/2$. Different values of $L$ are considered, as shown by the legend. The solid line scales like $L^3$.}
 \end{figure}

In general, the results are analogous to those obtained for regular structures and they do not depend qualitatively on the distribution\footnote{This has been checked for normal and uniform distributions and, more generally, it is expected to hold for a large class of distribution fulfilling the central limit theorem.} from which $L_y$ is drawn, hence conferring to the overall picture a great robustness.

In particular, here we show results obtained when the length of side-chains is extracted from a uniform distribution in the range $[1, \bar{L}]$, so that every integer $i$ in this range has the same probability, $1/\bar{L}$, to be chosen and the mean length is $\langle L_y \rangle = \sum_{i=1}^{\bar{L}} i/\bar{L} = \bar{L}(\bar{L}+1)/2$. Numerical results for the mean time $\tau_{\mathcal{B}}(L,\alpha)$ for two mobile particles starting from the same site on the backbone to meet again for the first time, 
the mean time $\tau^{\prime}_{\mathcal{B}}(L,\alpha)$ for a mobile particle to first return to the starting point on the backbone, 
the mean time $\zeta_{\mathcal{B}}(L,\alpha)$ for two mobile particles to first encounter having started on points in the backbone at a distance $L/2$, and
the mean time $\zeta_{\mathcal{B}}(L,\alpha)$ for a mobile particle to first reach a site at a distance $L/2$ on the backbone 
are shown in Fig.~$12$ (panel $a$, $b$, $c$, and $d$, respectively). The mean values obtained in this context have been calculated by averaging over both the underlying random structures and over different realizations of the two random walks; the latter sampling turns out to be more noisy than the former and it basically determines the final error to be associated to the mean time.

The behavior of the quantities mentioned above can be summarized as follows:
\begin{eqnarray}
\label{eq:tt1}
\tau_{\mathcal{B}} &\sim& \langle L_y \rangle^{\gamma} L_x\\
\tau^{\prime}_{\mathcal{B}}  &\sim& \langle L_y \rangle^{\gamma} L_x^{3-\gamma} \\
\zeta_{\mathcal{B}} &\sim& \langle L_y \rangle L_x  \\
\label{eq:tt2}
\zeta^{\prime}_{\mathcal{B}} &\sim& \langle L_y \rangle  L_x^2.
\end{eqnarray}


We notice that no fundamental difference emerges compared to the case of deterministic combs (see Eqs.~\ref{eq:tc}, \ref{eq:zeta}, \ref{eq:tauc2}, and \ref{eq:zetac2}, respectively). 
 
We also checked that these results are qualitatively robust with respect to the introduction of random ``defects'', such as the insertion of a small (i.e. sublinear with respect to $L$) number of links connecting nodes belonging to adjacent teeth (hence implying loops).
 

Therefore, the slowing down of two-particles reactions seems to derive from the high degree of inhomogeneity exhibited by such bundled structures, constructed by engrafting a branch on each vertex of a linear chain. Remarkably, branches do not have to be strictly separate (i.e. loops may be allowed) and, by taking as base graph another recurrent graph, analogous slowing down phenomena are expected (see e.g., \cite{CR-PRL1995}).

\begin{figure}[h] \label{fig:TAU3b}
\includegraphics[width=8.75cm]{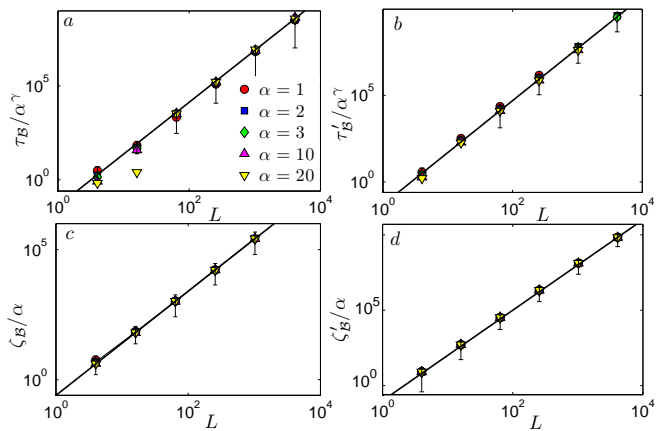}
 \caption{Normalized mean times for a particle starting from the backbone of a random comb $\mathcal{B}$ to first encounter another mobile particle with the same initial point (panel $a$), to first return to the initial point (panel $b$), to first encounter another mobile particle started at a distance $L/2$ (panel $c$), to first reach a site on the backbone at a distance $L/2$ (panel $d$). These mean times are divided by $\alpha$ to a proper exponent in order to obtain the data collapse; again, we took $\gamma=1.75$. Several choices of $\alpha$ are considered, as given by the legend in panel $a$. Solid lines represent the best fits which correspond to the following power laws: $\sim L^{1+\gamma}$, $\sim L^3$, $\sim L^2$, $\sim L^3$, respectively, hence suggesting the behaviors given in Eqs.~\ref{eq:tt1}-\ref{eq:tt2}. The data shown here have been obtained via numerical simulations; for every $L$ we extracted $10^2$ random structures and for each of them we considered a sample made of $10^5$ realizations. The error bars inserted in each panel refer to the data sets corresponding to $\alpha=1$ and are taken as representative for the whole ensemble of data. Notice that the error bars are asymmetric around the data points due to the logarithmic scale.}
 \end{figure}

\section{Discussion}
By explicitly studying specific examples we have shown that topological inhomogeneities deeply affect the kinetics of two particle encounter  processes even on finite structures. The main effect we evidenced is a strong slowing-down of the probability of encounter, compared with the situation for analogous regular structures.
In particular, it is possible to obtain transient kinetics, typical of higher dimensional structures, even in two-dimensional restricted geometries. This suggests a new strategy to control reaction kinetics: while, in order to increase the survival probability of a species, one usually increases the spatial dimension, by \emph{adding} sites, links  or volume to a given structure, in many cases it is possible to obtain a similar or stronger effect by judiciously \emph{deleting} elements, i.e. by sparing material instead of wasting it.
This opens the way to a new concept of geometrical tuning of chemical reactions, particularly suitable to restricted, low dimensional substrates.

\section*{Acknowledgements}
The FIRB grant RBFR08EKEV Sapienza Universit\`a di Roma, and GNFM are acknowledged for financial support.\\
Support of the DAAD through the PROCOPE program (project Nr. 55853833), of the European Community within the $7$-th Framework Program SPIDER (PIRSES-GA-2011-295302) and of the Fonds der Chemischen Industrie is acknowledged.



\bibliographystyle{ieeetr} 
\bibliography{RW}

\end{document}